\documentclass[twocolumn,showpacs,amsmath,amssymb, showkeys]{revtex4}

\usepackage{dcolumn}% Align table columns on decimal point
\usepackage{bm}% bold math\usepackage{amsmath}
\usepackage{graphicx}% Include figure files

\begin{document}

\title{Estimation of gravitational bending of light in the weak deflection limit}

\author {Arunava Bhadra}

\affiliation{High Energy $\&$ Cosmic Ray Research Centre, University of North Bengal, Siliguri, West Bengal, India 734013} 

\begin{abstract}

After discussing some subtle issues concerning the computation of deflection angle, a general but simple expression of bending angle of light rays in weak defelction limit has been presented for a general static and spherically symmetric space-time. In this context the importance of proper choice of polar axis has been highlighted. Applying the prescribed method the explicit expression of bending angle up to an accuracy of second order in mass is obtained for the Schwarzschild geometry without restricting the source and the observer to be at infinite distance away from the lens. 

\end{abstract}

\maketitle

\section{Introduction}

Light rays while passing a massive body are bent towards that mass. The prediction of general relativity in respect to the angle of deflection for the light ray just grazing the limb of the Sun was confirmed in 1919 when the apparent angular shift of stars close to the limb of the sun was measured during a total solar eclipse. This gravitational deflection angle has since been repeatedly confirmed with high-precision measurements and it provides compelling evidence in favor of general relativity. An important application of the deflection effect concerns the so-called gravitational lens that acts as a tool to study various features of the universe including estimation of masses of galaxies [1].  

Conventionally the quantum of bending of light rays due to a spherically symmetric mass distribution (lens) is computed by studying the trajectory of light rays in the gravitational field of the lens. Usually light trajectory is obtained from the null geodesic equation either using perturbative approach (see for instance [2-3]) or by directly integrating the null geodesic equation for the orbit equation of light [4-5]. When both the source and the observer are situated far far away from the lens in compare to the distance of closest approach of light rays from the lens, the angle between the two asymptotes of the light orbit is considered as the bending angle though such an equality is not apparent from the lensing diagram. More importantly the justification for such an equivalence has not commonly been discussed in the standard monographs of general relativity. As a result several issues concerning estimation of bending angle remain unclear to many and the equality of bending angle and the angle between the two asymptotes of the light trajectory has often been used in situations where it is not truly valid. The existence of different methods of estimating bending angle has further enhanced the perplexity of the problem. Consequently different expressions of bending angle for the same lensing problem, particularly whenever gravitational bending is considered in situations other than the classic case of infinetly distant observer and source, have been found in the literature that creats greater confusion. 

The purpose of the present article is to address such subtle issues concerning computation of bending angle. We will particularly highlight the importance of the proper choice of polar axis for framing the deflection scenario and calculating the deflection angle. Subsequently we will describe a straightforward and general approach of computing bending angle when the source and /or observer are located at a finite distance away from the lens. After providing a rigorous analytical framework of estimating bending angle for the general static spherically symmetric metric we will calculate the explicit expressions of bending angle for the Schwarzschild geometry. 

The organization of the article is follows. In the next section we would discuss the null geodesic equations for the general static spherically symmetric metric. In section 3 we would write down the orbit equation of light in the Schwarzschild geometry. The deflection angle will be computed in section 4 and finally we would conclude in section 5. 

\section{Equation of motion for photons}

We consider the general static and spherically symmetric spacetime which is given by (we use geometrized units \textit{i.e.} $G=1$, $c=1$) 

\begin{equation}
ds^{2}=-B(\bar{r})dt^{2}+ A(\bar{r}) d\bar{r}^{2} + \bar{r}^{2} D(\bar{r}) \left( d\theta ^{2}+ sin^{2}\theta d\phi ^{2}\right) 
\end{equation}

In principle one may recast the above metric to the standard form $ ds^{2}=-f(r)dt^{2}+ g(r) dr^{2} + r^{2} \left( d\theta ^{2}+ sin^{2}\theta d\phi ^{2}\right)$ just by transforming to a new radial variable $r$ where $r \equiv \bar{r} \sqrt{D(\bar{r})}$, but in some cases no explicit expression for the transformation equation exists or the transformation equation is too complex to deal with. Hence at this stage we would continue with the form as given by Eq. (1).     

Instead of obtaining the geodesic equations directly from the standard equations for a geodesic $\frac{d^{2}x^{\lambda }}{ds^{2}} + \Gamma _{\mu \nu }^{\lambda }\frac{dx^{\mu } }{ds}\frac{dx^{\nu }}{ds}=0 $, one may get those from the Euler$–$ Lagrange kind of equations, viz. [2]

\begin{equation}
\frac{d}{dp} \left( \frac{\partial \zeta}{d\dot{x}^{\mu} } \right) - \frac{\partial \zeta}{dx^{\mu} } = 0   
\end{equation}

where $\zeta \equiv g_{\mu\nu} d\dot{x}^{\mu} d\dot{x}^{\nu}$ is the square of the Lagrangian ($L$) of the gravitational field, $d\dot{x}^{\mu}\equiv dx^{\mu}/dp$, p is some affine parameter along the geodesic $x^{\mu} (p)$.

The square of the Lagrangian for the general metric (1) is given by

\begin{equation}
\zeta =-B(\bar{r})\dot{t}^{2}+ A(\bar{r}) \dot{\bar{r}}^{2} + \bar{r}^{2} D(\bar{r}) \left( \dot{\theta} ^{2}+ sin^{2}\theta d\dot{\phi} ^{2}\right) 
\end{equation}  

The resulting four geodesic equations are given by 

\begin{equation}
\frac{d}{dp} \left( B \frac{dt}{dp} \right) =0
\end{equation}%

\begin{eqnarray}
2 A \frac{d^{2}\bar{r} }{dp^{2}} + A^{^{\prime }} \left(\frac{d\bar{r} }{dp} \right) ^{2} +   B^{^{\prime }} \left( \frac{dt}{dp} \right) ^{2} - \nonumber \\
\left(D^{^\prime} \bar{r}^{2} + 2\bar{r}D \right) \left(\left( \frac{d\theta}{dp} \right) ^{2}+ sin^{2}\theta \left( \frac{d\phi}{dp} \right) ^{2}\right)
 &=& 0
\end{eqnarray}

\begin{equation}
\frac{d}{dp} \left(\bar{r}^{2} D \frac{d\theta }{dp} \right) - \bar{r}^{2} D sin\theta cos\theta \left( 
\frac{d\phi }{dp}\right) ^{2}=0
\end{equation}%

\begin{equation}
\frac{d}{dp} \left(\bar{r}^{2} D \frac{d\phi }{dp} \right) = 0
\end{equation}%

where primes denotes differentiation with respect to $\bar{r}$. If we choose $
\theta =\pi /2$ and $d\theta /dp=0$ initially, Eq.(6) warrants that they
would remain the same always. Thus for simplicity hereafter we would consider the motion only in the equatorial plan. 

The Eq.(4) straightway gives

\begin{equation}
\frac{dt}{dp}= \frac{k}{B} \;,
\end{equation}%

where $k$ is a constant. Normalizing time coordinate suitably we can set $k=1$. Integrating Eq.(7) 

\begin{equation}
\bar{r}^{2} D \frac{d\phi }{dp}= b
\end{equation}%

where $b$ is a constant of integration. The Eq.(5) together with Eqs. (8), (9) give

\begin{equation}
A \left( \frac{d\bar{r} }{dp} \right)^{2} - \frac{1}{B} + \frac{b^{2}}{ \bar{r}^{2} D} =  \frac{1}{E^{2}}
\end{equation}%

where $E$ is a constant of motion. Before proceeding further we would see what these constant of motion refer to. The Eqs. (1), (8), (9) and (10) together give $ds = dp/E$. For photon obviously $1/E =0$ whereas for massive particle $E>0$. Since $\zeta$ does not depend explicitly on $t$ and $\phi$, the $t$ and $\phi$ components ($p_{t}$ and $p_{\phi}$ ) of a particle's 4-momentum ($p^{\mu} \propto dx^{\mu}/ds$) are conserved along geodesics. For an observer at rest at infinity the energy per unit rest mass of a particle is $\epsilon=p_{t}$. Therefore, we have from Eq.(4) $\epsilon=g_{tt}p^{t}=E$. On the other hand $p_{\phi}=g_{\phi \phi} d\phi/ds =  \bar{r}^{2} D d\phi/ds = bE$. Thus $b$ represents the impact parameter at large distances. Here note that the unit of the constants of motion depends on the choice of afine parameter and the normalization of the time coordinate. For instance if one takes $dp \equiv ds$, then $b$ will be angular momentum per unit mass (or for mass less particle it is the angular momentum), $E=1$ and  $k$ (of Eq.(8)) would then describe the energy per unit rest mass. 

From Eqs.(8), (9) and (10), we finally obtain for photons

\begin{equation}
\frac{A}{\bar{r} ^{4}}\left( \frac{d\bar{r} }{d\phi }\right) ^{2} + \frac{D}{\bar{r}^{2}} - \frac{D^{2}}{B b^{2}} = 0
\end{equation}%

Integrating the above equation we will get $\phi$ as a function of $\bar{r}$ for the light orbit. However, to describe an orbit one usually expresses radial variable as a function of angular coordinate. Differentiating Eq.(11) with respect to $\phi$ one gets 

\begin{equation}
\frac{d^{2}u}{d\phi ^{2}} + 2\frac{D}{A} u + \frac{u^{2}}{2} \frac{d}{du} \left(\frac{D}{A} \right)  = \frac{1}{2b^{2}} \frac{d}{du} \left(\frac{D^{2}}{AB} \right)
\end{equation}
 
where $u=1/\bar{r}$. For a given metric i.e. for given $B$, $A$ and $D$ the solution of the above second order differential equation gives the photon orbit. 

\section{The orbit equation for photon in the Schwarzschild geometry}

The first step of calculating bending angle is to obtain the orbit equation. We would discuss hereunder both the popular approaches of solving geodesic equations for the light trajectory.

\subsection{Perturbation method}
 
Let us consider first the case when there is no lens and hence the geometry is flat (Euclidean) i.e. $A=B=D=1$. In such a situation the Eq.(12) reduces to 

\begin{equation}
\frac{d^{2}u}{d\phi^{2}}+u = 0 \;.
\end{equation}

Hence the orbit equation of undeflected (straight line) light ray in flat space not containing the origin of the coordinate system is given by 

\begin{equation}
u = \frac{1}{R} sin (\phi +\xi) \;,
\end{equation}

where $R$ is the perpendicular distance from the origin of the co-ordinate system to the path of the light rays and $\xi$ is the angle that the light rays made with the polar axis at the point of intersection. The angle $\phi$ is measured from the centre of the co-ordinate system with respect to the polar axis. Note that in the absence of any lens one is free to take any point as the origin of the coordinate system and also may set any direction of his/her choice as the direction of the polar axis.  

Now we consider the propagation of light ray in the gravitational field of the lens that describes a spherical distribution of mass. We are not considering the rotational motion of the lens/observer. Since we have to deal with the gravitational field of the Lens it is much convenient to take the Center of the Lens (L) as the origin of the coordinate system. 

To have a solution of the Eq.(12) one has to supply explicit expressions for $A$, $B$ and $D$. Now referring to the case of the Schwarzschild geometry  

\begin{equation}
ds^{2} = -(1-\frac{2m}{r})dt^{2} +  (1-\frac{2m}{r})^{-1}dr^{2} + 
r^{2}\left(d\theta^{2}+ sin^{2} \theta d\phi^{2} \right),
\end{equation}

i.e. when $B=1-2m/r$, $A=1/B$, $D=1$ and $\bar{r} =r$ the null geodesic equation (12) reduces to 

\begin{equation}
\frac{d^{2}u}{d\phi^{2}}+u=3mu^{2} \; .
\end{equation}

When the gravitational field is weak we can expect that the light path will deviate by a small amount from the undeflected (straight line) path and accordingly in such a situation we will look for a perturbative solution about the undeflected orbit. Therefore, we may take the solution as $u = u_{o} + u_{1}$, where $u_{o}$ is the undeflected trajectory as given by Eq. (14), and $u{1}$ is small perturbation. Substituting this form of $u$ along with the explicit expression of $u_{o}$ in to the Eq.(16) and neglecting higher order terms in $u_{1}$, one would obtain the expression for $u_{1}$ as follows

\begin{equation}
u_{1}= \frac{3m}{2R^{2}} \left[1+\frac{1}{3}cos 2(\phi + \xi) \right] 
\end{equation}

For the next higher order correction term $u_{2}$, one may follow exactly in the same way as before i.e. by writing $u = u_{o}+u_{1} + u_{2}$, substituting the expressions of $u_{o}$ and $u_{1}$ in to Eq.(16), ignoring higher order terms in $u_{2}$ and finally solving Eq.(16) for $u_{2}$. The resulting solution for photon orbit which upto the second order accuracy in $m$ is given by 

\begin{eqnarray}
1/r&=&u=\frac{1}{R} sin (\phi + \xi) + \frac{3m}{2R^{2}} \left[1+\frac{1}{3}cos 2(\phi + \xi) \right] \nonumber \\
&& + \frac{3m^{2}}{16R^{3}} [10\pi cos (\phi + \xi) - 20(\phi + \xi)cos(\phi 
 + \xi) \nonumber \\ 
&& -sin3(\phi + \xi) ]
\end{eqnarray}

The above equation represents the orbit equation for the light ray up to an accuracy of $m^{2}$. Since the shape of the orbit has changed due to the presence of the lens the distance of closest approach is also likely to be altered. Differentiating both sides of the above equation with respect to $\phi$ one gets  

\begin{eqnarray}
\frac{1}{r^{2}}\frac{dr}{d\phi} &=& - \frac{cos (\phi + \xi) }{R} \left[1 - \frac{2m}{R} sin (\phi + \xi) \right]  \nonumber \\
&& + \frac{15m^{2}}{4R^{3}}[cos (\phi + \xi)-\frac{3}{20}cos3(\phi + \xi) \nonumber \\
&& +(\pi/2-\phi-\xi)sin (\phi + \xi) ]
\end{eqnarray}

At the point of closest approach $\frac{dr}{d\phi}$ vanishes which occurs at $\phi + \xi = \pi/2$. Thus the distance of closest approach $r_{o}$ is given by

\begin{equation}
\frac{1}{r_{o}} = \frac{1}{R} \left(1 + \frac{m}{R} + \frac{3m^{2}}{16R^{2}}\right)  \;.
\end{equation}

Here it is to be noted that $r_{o}$ is smaller than $R$ (to the leading order $r_{o} \sim R -m$) i.e. in the presence of the lens the distance of closest approach decreases which is quite natural in view of the attractive nature of the lens. But in that case light rays starting from the source will not reach the observer as the source and the observer positions are fixed a priori. So we have to ensure that light path is through both the source and the observer. When such an aspect is taken into consideration the parameter $R$ in Eq. (18) will have to be larger than $R$ of the orbit equation for undeflected light rays (Eq. (14)). Accordingly in presence of the lens the distance of closest approach will become larger for the light trajectory between the source and the observer in compare to that of the undeflected light orbit (in the absence of the lens).   

The Eq.(18) has to be a solution of the parent Eq.(11) with $B=1-2m/r$, $A=1/B$, $D=1$. This requirement implies $b=R(1+m/R)$ to the leading order in m. The Eqs.(18) and (20) completely describe the photon orbit. In the small angle approximation the photon orbit can be expressed in terms of the physically meaningful variable $r_{o}$  as

\begin{equation}
1/r = u = \frac{\phi + \xi}{r_{o}} + \frac{2m}{r_{o}^{2}} + 
  \frac{m^{2}}{8r_{o}^{2}} \left(15 \pi -16\right)  + {\cal O}\left(m^{3} \right)
\end{equation}
 
\subsection{Integration method} 

Integrating the equation (11) one gets

\begin{equation}
\phi = \pm \int \frac{ A^{\frac{1}{2}} d\bar{r} }{\bar{r} ^{2}}  \left[ \frac{D^{2}}{B b^{2}}  -\frac{D}{\bar{r} ^{2}}\right] ^{-\frac{1}{2}}\;
\end{equation}

which is also the equation of light orbit. As stated before one has to provide the metric coefficients $A$, $B$ and $D$ for explicit expression of photon trajectory. However, for most of the relevant cases the above integration cannot be done analytically with any order of accuracy. Hence one usually expands $A$, $B$ and $D$ in terms of $m/\bar{r}$, retain terms upto a desired order of accuracy and then integrate the above equation to obtain photon orbit [4-5]. 

To exemplify let us consider the case of the Schwarzschild metric. In this case the Eq.(22) is integrable and the exact solution of $\phi$ can be described in terms of incomplete elliptic integral of first kind [6] but such a form of the solution is not very illuminating, particularly for practical purposes. Instead usually the metric coefficients are approximated up to a certain order to get the approximate deflection angle. Here we are looking for a solution upto an accuracy of second order in $m/r$. Computing light trajectory to any given order requires the knowledge of every terms of the metric coefficients to that order in the expansion [7]. Hence in the present case our metric coefficients will be $B=1-2m/r+2m^{2}/r^{2}$, $A=1+2m/r + 3m^{2}/(2r^{2})$ and $D=1$. From the fact that at the point of closest approach $dr/d\phi$ must vanish, we get from the Eq.(11)  
\begin{equation}
b=r_{o}\left(1+ \frac{m}{r_{o}} + \frac{3}{2} \frac{m^{2}}{r_{o}^{2}}\right) + {\cal O}\left(m^{3} \right)
\end{equation}

Inserting the expression of $b$ in to the Eq. (12) and integrating it we get   

\begin{eqnarray}
\phi (r) &=& \xi_{o} - tan^{-1}\frac{r_{o}}{ \sqrt{ r^{2}-r_{o}^{2} } } + \frac{m \sqrt{ r^{2}-r_{o}^{2} }}{r_{o}} \left(\frac{1}{r} + \frac{1}{r+r_{o}} \right)  \nonumber \\ 
&& + \frac{3m^{2}}{4} \left(\frac{\sqrt{r^{2}-r_{o}^{2}}}{r_{o} r^{2}} - \frac{1}{r_{o}^{2}} tan^{-1}\frac{r_{o}}{\sqrt{r^{2}-r_{o}^{2}}} \right)  - \nonumber \\
&&   -\frac{3m^{2}}{r_{o}^{2}} \left(\sqrt{\frac{r-r_{o}}{r+r_{o}}} + tan^{-1}\frac{r_{o}}{\sqrt{r^{2}-r_{o}^{2}}} \right)   - \nonumber \\
&& \frac{m^{2} \sqrt{r^{2}-r_{o}^{2}} (2r + r_{o})}{2r_{o}^{2} (r+r_{o})^{2}} + {\cal O}\left(m^{3} \right)
\end{eqnarray}

where $\xi_{o}$ is an integration constant. When $r >> r_{o}$, the above solution becomes

\begin{equation}
\phi (r \rightarrow \infty) = \xi_{o} + \frac{2m}{r_{o}} - \frac{2m^{2}}{r_{o}^{2}}
\end{equation}

Note that both the Eqs. (18) and (24) describe the same photon trajectory; in the former expression $r$ has been expressed in terms of $\phi$ while the reverse has been done in the later equation (though one may not obtain the one from the other easily). 

\section{Deflection angle}

In the Schwarzschild geometry in order not to be captured, the apparent impact parameter $b$ has to be at least greater than the critical value $b_{c} \equiv 3\sqrt{3}m$ [4]. Here we are considering deflection in the weak gravity scenario. So in our case $b$ is quite larger than the critical value $b_{c}$.   

The geometrical configuration for the phenomenon of gravitational bending of light in asymptotically flat space time is portrayed in Figure 1. The light emitted by the distant source S is deviated by the gravitational source (Lens) L and reaches the observer O. If both the observer and the source are situated in the flat spacetime region then the angle of deflection ($\alpha \equiv \angle PQO$) by the lens is the difference between the angle of emission (from the source) and the angle of reception (by the observer) with respect to a common polar axis. 

\begin{figure}
\centering
\includegraphics[width=0.75\textwidth,clip]{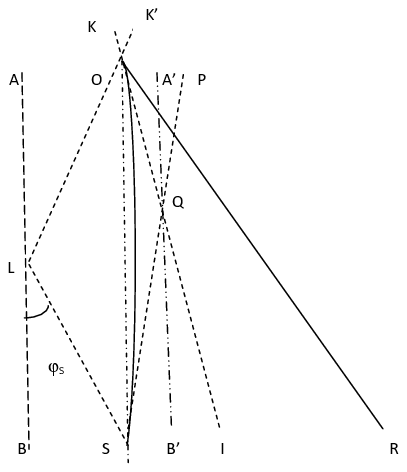} \hfill
\caption{The geometrical configuration of gravitational bending of light.}
\end{figure}

The difference in angular coordinates of the source ($\phi_{S}$) and the observer ($\phi_{O}$) is often ascribed in the literature (see for instance [2-5]) as the bending angle ($\delta \phi \equiv \phi_{S}(\infty) - \phi_{O} (\infty)$) when the source and the observer are at infinitely large distances away from the lens. But how $\delta \phi$ becomes equal to the bending angle under the stated limit on the source/observer distance is not very instructive from the Figure 1. The justification for such equivalence is discussed below. 

In this respect the first crucial aspect is to set the polar axis. Note that the center of the lens has already been set as the origin of the co-ordinate system but so far no restriction has been imposed on the orientation of the polar axis. One is thus free to take any line passes through the point L as the polar axis. But the conventional strategy of determining bending angle does not work for an arbitrary orientation of the polar axis. For instance consider that the optic axis i.e. the line joining the observer and the lens is the polar axis (OL). Accordingly $\phi_{O}=0$ irrespective of the distance of the observer from the lens whereas $\phi_{S}$ will be a non-zero constant unless the particular case of perfect alignment between the source, the lens and the observer. Once $\phi_{S}$ is fixed, the $\delta \phi$ i.e. the angular difference between the source and the observer would not change with the change in source distance and more importantly this angle certainly does not give the bending angle. In such a choice of the polar axis the difference of offset angles $\xi$ at the location of the source and the observer should directly give the bending angle but there is no simple way to determine these angles.  

The standard methods require that the polar axis to be parallel (perpendicular) to the undeflected light ray (AB in Fig.1). In that case the offset angle $\xi $ of Eqs. (14), (17), (18) , (19) becomes zero ($\pi/2$). As $r \rightarrow \infty$, $\phi $ becomes very small. Denoting $\phi = \phi (\infty)$ (or $\pi - \phi(\infty)$) when $r \rightarrow \infty$ we get from the Eq.(18) that $\phi(\infty) = -2m/R -\frac{15\pi m^{2}}{8R^{2}}$ (or $\pi + 2m/R + \frac{15\pi m^{2}}{8R^{2}}$). The lensing geometry emerged from these results is shown in Fig.2. Since the observer is at an infinte distance away from the lens, the light path LO will be parallel to OQ. Hence the angle $\phi_{O} (\infty) = \angle OLA = \angle OQA^{\prime}$. Similarly for the source $\phi_{S} (\infty) = \angle SLB = \angle SQB^{\prime} $. Hence the resulting deflection angle $\angle PQO = \angle A^{\prime}QP + \angle B^{\prime}QI = -4m/R$ which is the standard result. The computation of bending angle using the direct integration method is also based on the same principle. 

\begin{figure}
\centering
\includegraphics[width=0.5\textwidth,clip]{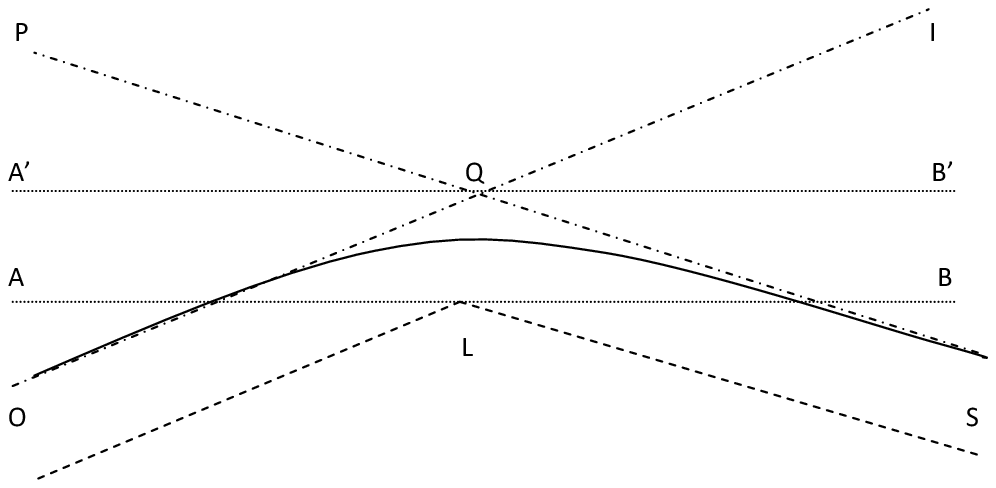} \hfill
\caption{Gravitational bending geometry when the source (S) and the observer (O) are at infitely large distance away from the lens (L). AB is the polar axis, $A^{\prime}B^{\prime}$ is parallel to the polar axis passing through the point Q, I is the image position, R is a reference object}
\end{figure}

When the source and/or the observer are located at a finite distance away from the lens, $\delta \phi$ no longer does represent the deflection angle. It just gives the angular separation between the source and the observer. This separation of angle is independent of the light path direction and should be the same for deflected and undeflected light rays. Let us go back to the Figure 1. The line $A^{\prime}B^{\prime}$ is parallel to AB passing through Q. So deflection angle $\angle PQO = \angle A^{\prime}QO + \angle B^{\prime}QS$. $\angle A^{\prime}QO = \angle KOK^{\prime} - \angle ALO$. So if $\psi_{O}$ describes the angle that the tangent of the light trajectory made with at the position of observer with the line $\phi_{O} = constant$, where $\phi_{O}$ denotes the angular coordinate of the observer then $\angle A^{\prime}QO = \psi_{O} - \phi_{O}$. Similarly $\angle B^{\prime}QS = \psi_{S} - \phi_{S}$. So the total bending angle  $\angle PQO = \psi_{O} + \psi_{S} - \phi_{O} - \phi_{S}$.   

The invariant formula for the cosine of the angle between two coordinate directions P and Q 

\begin{equation}
cos(\psi) = \frac{ g_{ij} P^{i} Q^{j} } { (g_{ij} P^{i} P^{j})^{1/2} (g_{ij} Q^{i} Q^{j})^{1/2} }
\end{equation} 

If $P$ is taken as the direction of the orbit and $Q$ is taken as that of the coordinate line $\phi $ constant, then one may write $P \equiv (dr, d\phi) = (dr/d\phi,1)d\phi$ ($d\phi lt 0)$ and $Q \equiv (dr, 0) = (1,0)dr$. Consequently for the general spherically symmetric space time metric given by Eq. (1) the angle between $P$ and $Q$ directions becomes [2,8] 

\begin{equation}
tan(\psi) = \frac{\bar{r}\sqrt{D}}{\sqrt{A}} |d\phi/d\bar{r}|
\end{equation} 

The Eqs. (11) and (23) straightway give

\begin{equation}
d\phi/d\bar{r} = \frac{\sqrt{A(\bar{r})}}{\bar{r} \sqrt{D(\bar{r})}} \left[\frac{D(\bar{r}) B(r_{o})}{D(r_{o}) B(\bar{r})} \frac{\bar{r}^{2}}{r_{o}^{2}} -1\right]^{-1/2}
\end{equation}

Substituting in to Eq.(26), we get

\begin{equation}
tan (\psi) = \left[\frac{D(\bar{r}) B(r_{o})}{D(r_{o}) B(\bar{r})} \frac{\bar{r}^{2}}{r_{o}^{2}} -1\right]^{-1/2}
\end{equation} 

So once the metric is given, one can readily obtained the angle $\psi$ at any point on the light trajectory using the above expression. This is one of the main results of the present work.

The angle that the light trajectory makes with the direction of the polar axis is then given by (assuming the gravity is very weak at the positions of the observer/source)  

\begin{equation}
\epsilon_{i} = \psi_{i} - \phi_{i} \;, 
\end{equation} 

where the subscript $i$ denotes the source, observer. The bending angle ($\alpha$) will be then 

\begin{equation}
\alpha = \epsilon_{S} + \epsilon_{O}. 
\end{equation} 

For $\epsilon$, however, the knowledge of $\phi_{i}$ are needed that can be obtained from the light trajectory. So for a given metric employing the Eqs.(26) and (27) one would obtain the bending angle. 

When applied to the Schwarzschild metric, we get to the leading order in $m$ and $1/r^{i}$

\begin{equation}
tan (\psi_{i}) = \frac{r_{o}}{r_{i}} + \frac{m}{r_{i}}  -\frac{m r_{o}}{r_{i}^{2}}
\end{equation} 

Exploiting Eqs.(18) and (20) we finally get for small $\phi_{i}$

\begin{equation}
\epsilon_{i} = \frac{2m}{r_{o}} + \frac{m^{2}}{8r_{o}^{2}} \left(15\pi -16\right)  -\frac{m r_{o}}{r_{i}^{2}}
\end{equation} 

One may also determine the above expression using Eq.(24) for which one has to compute the change in angle between the source/observer and the point of closest approach ($r_{o}$). 

Therefore the resulting deflection angle becomes

\begin{equation}
\alpha = \frac{4m}{r_{o}} + \frac{m^{2}}{4r_{o}^{2}} \left(15\pi -16\right) -m r_{o} \left( \frac{1}{d_{OL}^{2}} + \frac{1}{d_{LS}^{2}} \right) 
\end{equation} 

where $d_{OL}$ and $d_{LS}$ are the radial coordinate of the observer and the source respectively. The above expression agrees with the previous results [9-13], particularly when $d_{OL}$ and $d_{LS}$ tend to $\infty$. However it is worthwhile to note that for finite $d_{OL}$ and $d_{LS}$ the above expression has one or two extra terms in compare to many of the earlier expression of the bending angle. For example, in [10-13] the third term of the right hand side of the above expression is missing which can contribute significantly in the solar system observations when one is looking for an accuracy of second order or better. Another point to be noted is that apparently at the second order level the expression given in [9] differs from the above expression. This is because in [9] isotropic coordinates ($t,\rho,\theta, \phi$) was used instead of the standard (Schwarzschild) coordinates ($t,r,\theta, \phi$) as used in this work. Hence the stated difference can be accounted by applying the transformation equation between isotropic and standard coordinates, namely $r=\rho (1+m/2\rho)^{2}$.

Since $r_{o}$ is a coordinate dependent quantity, it is not a measurable one and hence it is better to express the bending angle in terms of a proper quantity such as the apparent impact parameter $b$ exploiting Eq.(23). Accordingly, the deflection angle is given by

\begin{equation}
\alpha = \frac{4m}{b} + \frac{15\pi}{4} \frac{m^{2}}{b^{2}} -m b \left( \frac{1}{d_{OL}^{2}} + \frac{1}{d_{LS}^{2}} \right) 
\end{equation}    

The present result contains a new term involving distance of the source (last term of the right hand side of the above equation) in compare to the expression given in [9]. When the source will be a nearby one, such as in the planned astrometric missions using optical interferometry [14], this term will also contribute significantly. The expression of bending angle in [9] incorporates other effects, such as contribution of the angular momentum and quadrupole moment of the lens on the bending angle, which are not considered here.

\section{Discussion}

In the present work we have discussed the basic method of computation of gravitational bending angle and related subtle issues to clear the prevailing confusions on the topic. We stressed the importance of proper choice of polar axis in estimating the bending angle. After developing a general and rigorous analytical framework for computation of bending angle we presented a simple but general expression for computing bending angle without restricting the distances of the source and the observer at infinitely large level. Applying the method to the Schwarzschild geometry we obtained the explicit expression of the bending angle up to an accuracy of $m^{2}$. 

In estimating the bending angle one usually works in a particular coordinate system and finally the bending angle is presented as a function of coordinate distance of closest approach. For instance, if one computes the deflection angle in the standard coordinates, the bending angle is expressed in terms of $r_{o}$ whereas working with isotropic coordinate the bending angle is given as a function of $\rho_{o}$. A coordinate length like $r_{o}$ or $\rho_{o}$ is not directly measurable, they can only can be indirectly constructed from the values of actual measurements. Identifying the measured radius of an object with  coordinate distance of closest approach works tolerably well only up to the first order level but such an approximation does not work at second or higher order in $m$ [7]. So expressing bending angle in terms of coordinate distance of closest approach is not a good idea at all. One may instead express as a function of $b$ since the apparent impact parameter $b$ is a coordinate independent quantity. But in practical purpose $b$ is also not suiable as it is not possible to measure in a real experiment. It appears that the study of bending of a massive particle offers the only way of testing gravity at the second order level [7] through gravitational deflection measurements.        

Recently there is a serge of interest in the possibility of measuring the cosmological constant through its effect on the deflection of light after Rindler and Ishak [8] demonstrated that cosmological constant contributes to the bending angle against the conventional concept of non-contribution. However, the expressions of bending angle in the presence of cosmological constant obtained by different authors found to differ significantly [15]. The method prescribed in this work has found useful application in this respect [16]. The present approach is also expected to have useful implication to the lensing phenomenon in the strong field deflection limit [17] where the magnitude of bending is very small. A work has been undertaken in this direction.

\end{document}